# Educational Content Management – A Cellular Approach

*M. Engelhardt, A. Hildebrand, A. Kárpáti, T. Rack, T.C. Schmidt*

Fachhochschule für Technik und Wirtschaft Berlin

**Abstract:**

*In recent times online educational applications more and more are requested to provide self-consistent learning offers for students at the university level. Consequently they need to cope with the wide range of complexity and interrelations university course teaching brings along. An urgent need to overcome simplistically linked HTML content pages becomes apparent.*

*In the present paper we discuss a schematic concept of educational content construction from information cells and introduce its implementation on the storage and runtime layer. Starting from cells content is annotated according to didactic needs, structured for dynamic arrangement, dynamically decorated with hyperlinks and, as all works are based on XML, open to any presentation layer. Data can be variably accessed through URIs built on semantic path-names and edited via an adaptive authoring toolbox.*

*Our content management approach is based on the more general Multimedia Information Repository MIR. and allows for personalisation, as well. MIR is an open system supporting the standards XML, Corba and JNDI.*

## 1   Introduction

The quality of online learning applications mainly depends on two ingredients: The content itself and its presentation, the latter including hypermedia elements of interactivity. Today's ease in presenting multimedia and hypermedia content on the net tends to have us disregard that casually designed learning material in hypermedia appears particularly incoherent, nonsignificant and disappointing to the student [1].

Well prepared and maintained electronic content and its management in ODL faces a diversity of extra demands, among them

- coherence and timeliness of information
- reuse of simple, compound and fragmented content material
- dynamic content structuring and rearrangement with coherent presentation
- ease in authoring and updating of content constituents
- flexible options of content decoration with meta-data
- content retrieval and access based on semantics.

In addition particular attention needs to be drawn to the way linking is done within an application. Since "simply linking one text to another fails to achieve the expected benefit of hypermedia and can even alienate the user" (G.P. Landow, [1]) a coherent, transparent rhetoric scheme of setting hypermedia links within content units should be applied.

Any of those demands cannot be easily achieved or reached at all manually and should be supported by an educational content management system. Most of the above features, however, remain unseen in current learning environments. The primary, most often violated fundamental principle for educational content management we see in the strict separation of structure, logic, content and design, as it can be achieved by applying XML-technologies in a rigorous fashion. Here it should be noted that hyperlinks in our view belong to structural information and therefore must not be stored within content.

The view of the content presented to the students should be seen as part of the didactical model. The generation of views thereby in general include the provision of navigational and link structuring, of rhetorical and narrative devices as the result of dynamical (meta-) content processing, personalisation capabilities and the system ability to adapt to personal requirements. As they are elements of the didactical concepts all properties of the views need to remain subject of flexible configuration and modelling. Flexibility though ends where the author is driven towards the revocation of presentational concepts, as it is likely for manually encoded pages. The modelling therefore should be located on a separate layer, effective to the entire application, thereby ensuring a coherent view to the learner.

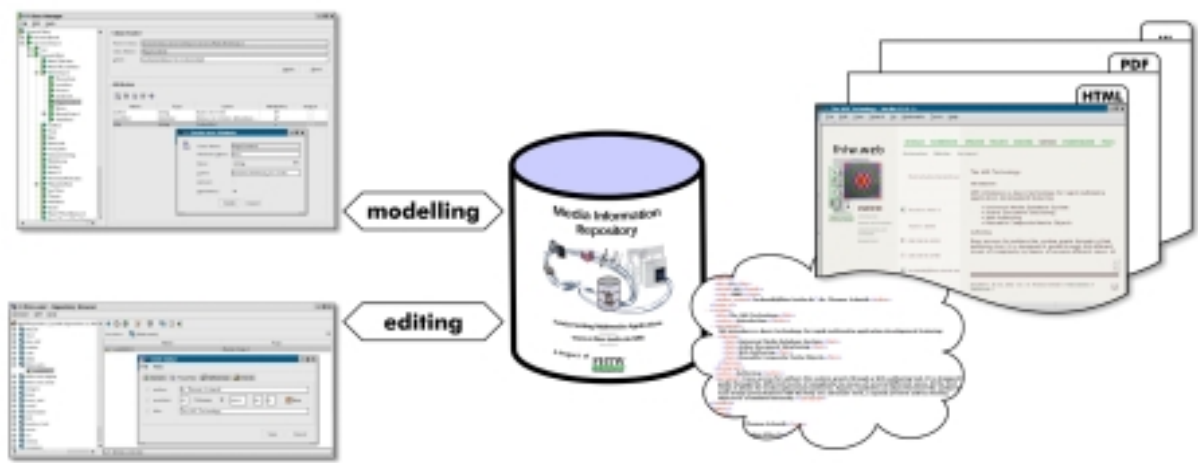

Figure 1: MIR provides flexible adaptive information structures

This report concentrates on our project activities of modelling and implementation of an educational content management system, which supports most of the above mentioned requirements. Our work covers concepts and implementation of organising and retrieving content and its meta-data, modelling static and dynamic (hyper-) structures, authoring and viewing content pages in a context sensitive fashion. Our immediate practical application will be to manage our universities website fhtw.web. More intricate learning applications are on schedule.

The field of educational hypermedia systems, though quite old, continues to show very active research and development activities. Numerous concepts, technologies and platforms presently are under work or design, the most prominent technological framework being XML-related [2]. Our work ranks around XML formats and technologies, as well, and relies on the more general storage and runtime platform Multimedia Information Repository (MIR) [3]. Built on a three-tiered architecture MIR provides all fundamental support of media data handling, authentication, user and connection handling. MIR is built as an open system and currently supports the standards XML, Corba and JNDI.

Grounded on a powerful media object model MIR was designed as a universal fundament for an easy design of complex multimedia applications. By means of its fully object oriented design MIR attains parametrisable, flexible data structures, thereby offers reusability of content objects at any level of complexity. It allows for modelling of almost arbitrary content entities with the aid of an object class modelling tool (see fig.1). A generic, only type dependent editing of data is part of the system, as well. MIR provides two layers of structuring content components, both open to application semantics: A passive referencing interrelates any entities according to static application structures. In addition a central benefit of our object oriented information model is the notion of active references as a basic composition mechanism. These active interrelations not only carry the ability to refer to subordinate presentation data, but are capable of imposing event-type actions on its references. For further reading we refer the reader to [4] and [5].

This paper is organised as follows. In section 2 we introduce basic concepts and solutions of content organisation and access. Section 3 presents the authoring concepts and tools. Section 4 is dedicated to describe MIRacle, our new ansatz in hyperreferencial interactivity. Finally, section 4 gives a brief conclusion and an outlook on the ongoing work.

## 2   Building Content from Cells

### 2.1   Hypermedia Content Organisation

Hypermedia learning systems brought new challenges to the production and organisation of content. Oriented at a page view of limited dimension, nonlinearly ordered and under the request for changing relations and dynamical rearrangements the classical paradigms of illustrated written text ceases to hold. Even in the early days of hypermedia major aspects of a new approach to content management had been noted and denoted in the Dexter Hypertext Reference Model [6]. In 1988 at the Dexter Inn an abstract notion of Storage Layer was created, consisting of linked, possibly composite Components. In our further discussion we will stick to the terminology of the Dexter Reference Model.

Hypermedia content which is subject to automated processing requests for strong structuring which can be essentially achieved in two ways: Information material may be on the one hand decomposed on the component layer into many small entities. A text, for example, could be split over many files. On the other hand structuring can take place on the within-component layer by means of a sub-addressing scheme. The text within one file, for example, may be built according to a DOM tree. The major difference of the two approaches is reflected in the data access. Either a file system resp. database has to be searched for appropriate content constituents, or a mime-specific retrieval of fragments has to be performed within data units. The latter could be of Xpointer-type in text/XML documents, a frame addressing in video/audio data, a polygonal geometry allocation for images, … For a more detailed discussion on fragmenting see [8].

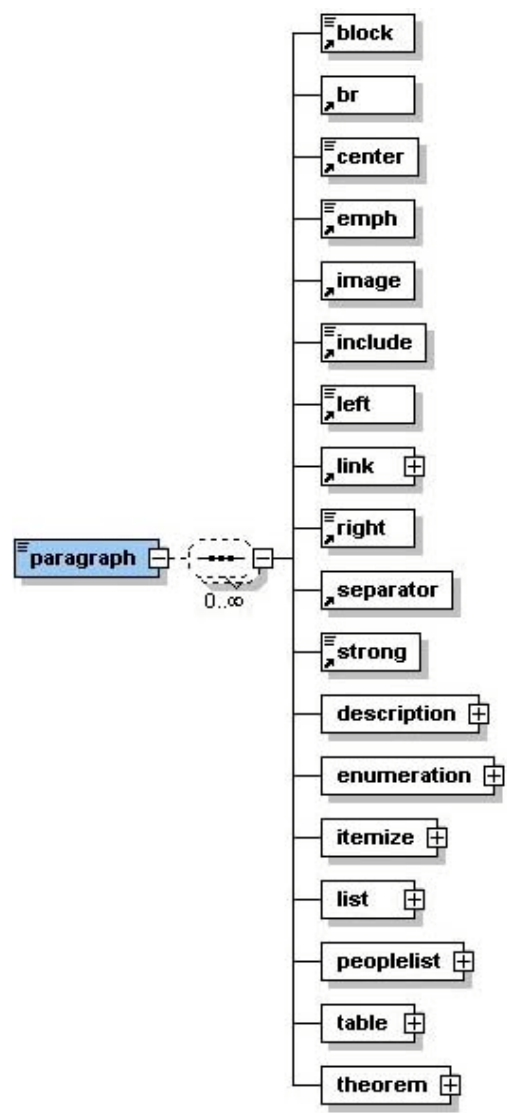

Figure 2: XML schema for paragraphs

Fragment addressing is much more complex and, from a computational point of view, expensive operation than component retrieval. On the other hand viewing and authoring of information entities consisting of many components is rather complicated. Being aware of these two extremes our cellular ansatz of content organisation forms a compromise: Text content of our solution is composite of cells, where cells are addressable elements consisting of an unstructured word at minimal and a text paragraph at most. The paragraph itself may be substructured according to the XML schema shown in figure 2.

In detail our content modelling proceeds as follows: All meta-data and particular content entities s.a. titles, authors, keywords, etc. are singled out as are directory-like information like teachers, course listings etc. Keeping these entities separate not only reaches for a high level of content normalisation, but also easily permits automatic generation of navigational overview, slides, etc., as well as updating content from the background. All other content units are organised in paragraphs which are collected to pages by means of the external structuring. Note that this concept implies that pages editing mainly consists of arranging paragraphs. In this way information entities of paragraph dimension are easily re-used by applying multiple structural references in a static or dynamic way. Note also that meta-annotation may be undertaken paragraph-wise as might be necessary to encode microscopic didactical concepts.

## *2.2 The Requirement for Context*

When formulating an abstract, flexibly meshed storage layer the Dexter group made a fundamental conceptional mistake. As was pointed out in the Amsterdam Hypermedia Model [7] the concept of composite components is incomplete without the notion of the context of components. When following a hyperlink pointing at a member of one or several compositions processing needs to consider the relevant composition, the actual context, in order to decide on actions to be taken.

Context is an inevitable part of the information needed for processing hypermedia. In HTML it is implicitly encoded via the embedding of hypermedia structure, i.e. links and anchors, within complete pages. In general there are two context situations to consider: The context of source or departure and the context of destination or arrival. In formal terms the context of a component is denoted by the data immediately embedding it. The notion of context generalises to fragments, as well [8].

While building content from composite components context needs some kind of extra encoding. This could be done by maintaining an additional information layer as is the 'perspective' approach in the Nested Context Model of Soares et. al. [9]. This fairly general concept suffers from the drawback of carrying an additional, partly redundant data structure. In [7] it has been already noted that content structures are suitable for carrying context information. Also to incorporate temporal information additional context nodes are invented by the Amsterdam group and contexts extracted from structure traversal. Note that context cannot be expressed by positioning components within a hierarchical file system, as it is often attempted in traditional webserver organisation: Object reuse and non-hierarchical component relations contradict a one-to-one correspondence of application contexts and a file system view.

However, context is implicitly encoded within the composite structure of the application. Accessing a component along the way of its currently valid composite references is equivalent to reconstructing its actual context. Starting from this observation to express contexts we introduce the concept of *context-sensitive* or *semantic paths*. Semantic paths are built of named relations within the composite structure, which can be seen as directories, and an appropriate access and retrieval logic. Applying these semantic paths the content appears organised in a file system like fashion, where the file system is neither hierarchical nor normalised, but directly reflects data contexts. Possible recursions thereby need to be treated by the access logic. Figure 3 schematically visualises the different views attained within the MIR database system. Whereas content components, Media Objects (MOBs) and Data Objects (DOBs), physically reside in a flat object store, they can be simultaneously accessed in a traditional, hierarchical virtual file system and in the semantic, context sensitive path space. Seen from the supported JNDI interface semantic paths form simply an alternate name space. Note that this semantic name space can be easily inverted to visualise the relation of 'Who references this component?'.

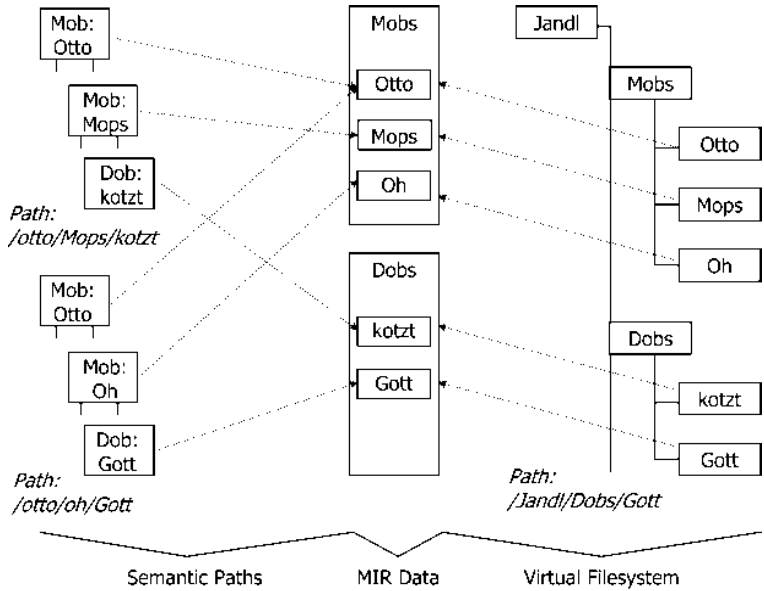

Figure 3: Semantic and hierarchical data view

In the framework of web-based information technology a vital interest lies in the provision of universal URI addressing. Any address should exhibit an URI representation. As our semantic paths add a new address space and as fragment elements within components should be URI-addressable, as well, the URI should cope with a four-layered hierarchy in address spaces (see fig. 4). Currently we work around this task by masking the global path with a virtual hostname, but in general a switch in namespace has to be marked. Thus context paths could be initiated with '#' and fragment search could find its (Xpointer-)expression following the '?'.

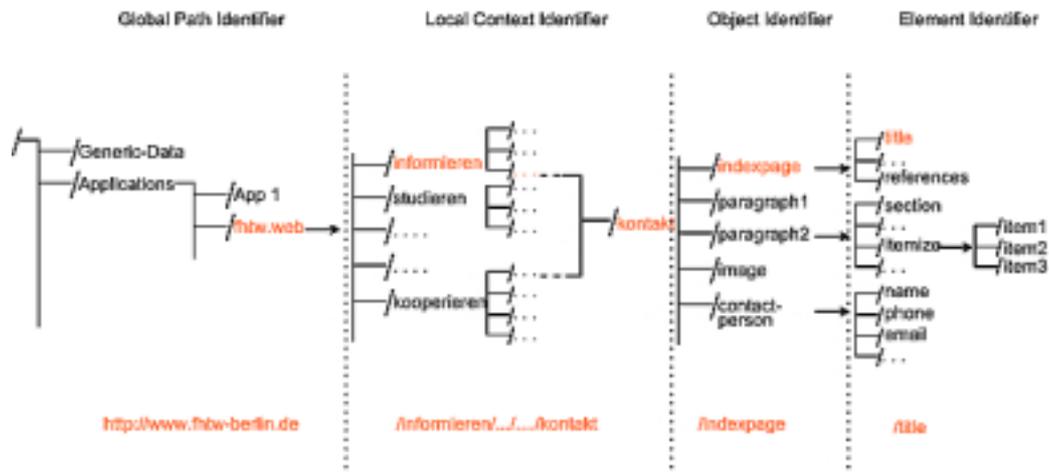

Figure 4: Address hierarchy for content access

# 3 Authoring within Contexts

Content forms the valuable heart of any educational application. Therefore great care should to be taken in providing appropriate tools for authoring and maintaining the information. Our cellular approach in content organisation on the one hand gives rise to particular ease in authoring the highly structured and coherently normalised material: Updating one information cell may lead to an actualisation of many views in several contexts. Many data cells, especially directory-like entries, are well suited for automatic update procedures. Cells on the other hand are by themthelves bare of context. Editing singular paragraphs on their own is inconvenient and is likely to disturb the author's train of thought.

Authoring continuous text via formulae is an even greater nuisance. It can be neither appreciated to build up larger documents with the browser's tool of HTML forms, nor is the XML editing likely to be accepted on the basis of formulae structured according to Schemas, as is the common approach of currently available XML-editors s.a. XML Spy. To tackle the authoring problem we designed an editor toolbox on the basis of JAVA/Swing, which dynamically adapts to the specific formatting requirements of content components. A WYSIWYG XML word processor for editing paragraphs is part of the tool set. As offering a WYSIWYG MS-Word-like editor for writing continuous text appears to be the only approach widely acceptable to the user, we mapped the structural elements of XML to common layout elements on the screen, thereby simulating an average type stylesheet. Since XML structuring is by no means congruent to formatting of text, this is a conceptual incorrectness, which we try to alleviate by displaying the structural entity on mouse-over.

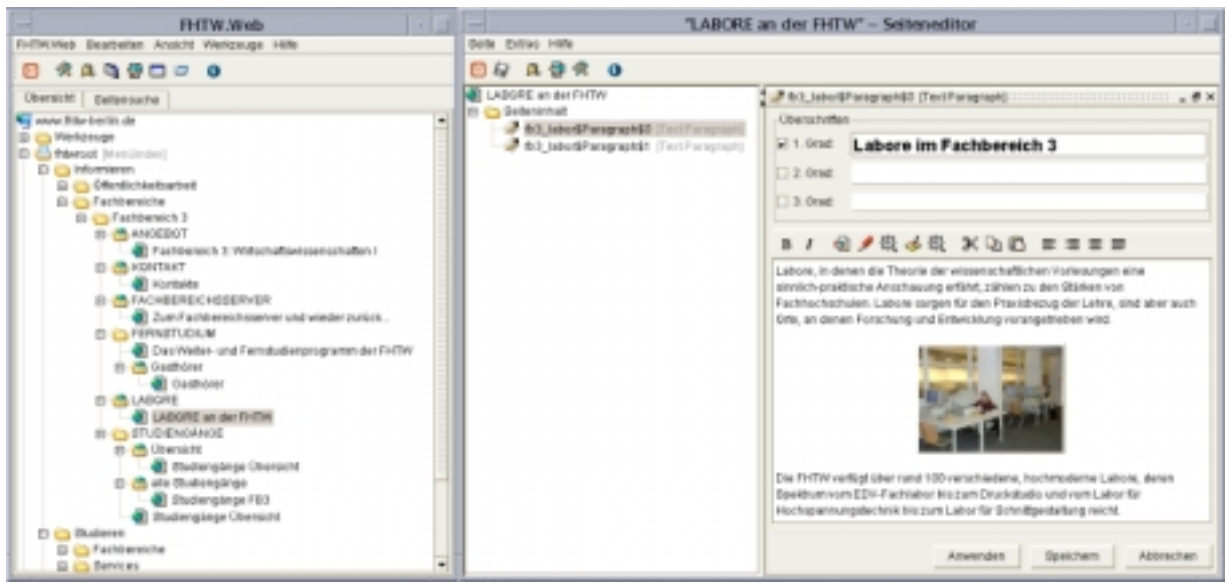

Figure 5: Authoring composite pages

The more challenging aspect of designing the authoring tool, though, is to present to the author an application specific view detached from specific content organisation. As content in online systems is presented as pages, pages should be shown for editing. Pages in our approach are a collection of cells arranged according to their page context. Pages can be addressed by navigating along the semantic path space explained in section 2 and edited by simultaneously opening its cellular components. The authoring as shown in figure 5 works that way. On the left the navigation according to the semantic file system is displayed, next to page editor with a specific page loaded. The editor itself provides a navigation through the relevant content structure of the page and a typing area combining related content cells for coherent editing.

# 4 MIRaCLE – an Adaptive Linking Environment

## 4.1 How to know the Right Way of Linking?

Interactivity, besides content, plays the second role to be stressed in educational learning systems. Well organised content can be significantly enriched or disturbed by the way links offering interactivity are added to it. As we already mentioned in the introduction do we not consider the definition of links as part of the content itself, but rather as part of the didactical structuring and presentation model. In particular, there should be a way to apply several linking schemes in different views to the same content.

To illustrate this argument consider the following example: A short introductory overview on hamster diseases written in the Gaelic language is presented to a Schottisch veterinarian who has significant weaknesses in Gaelic. The utmost help to him will probably be a linking, where every word is linked the corresponding one in the dictionary. An Irish hamster farmer with some semi-expert knowledge about his animals instead would profit most from having the medical terms linked to some dictionary entries. An Irish student of vet learning for his exam instead could appreciate all disease names being linked to some encyclopaedically knowledge for experts. And so on.

Thinking over this example we can extract the following requirements for an appropriate linking environment:

- There should be the option of applying different linking schemes to one content; thus the definition of links cannot be part of the content itself.
- Linking should adapt to the users requirements.
- A user should be enabled to adapt the linking of an application to his requirements.
- A rhetoric of linking [1] should be present and transparent to the user.
- High-level mechanisms for defining links are needed in order to keep work of the author simple.

Some of the above requirements can be tackled with the help of Xlink/Xpath/Xpointer [2], but major issues remain unsolved.

## 4.2 Concepts for an Adaptive, Dynamical Linking

The MIR adaptive Context Linking Environment was designed to meet all the above requirements. MIRaCLE is an adaptive scheme for dynamic link generation in internet-based teaching. The prototypically implemented model aims to enable teachers and students to semantically define linking behaviour. Here we give only a rough outline of the concept and the underlying ideas.

For a detailed discussion we refer to our forthcoming paper [10]. Abilities

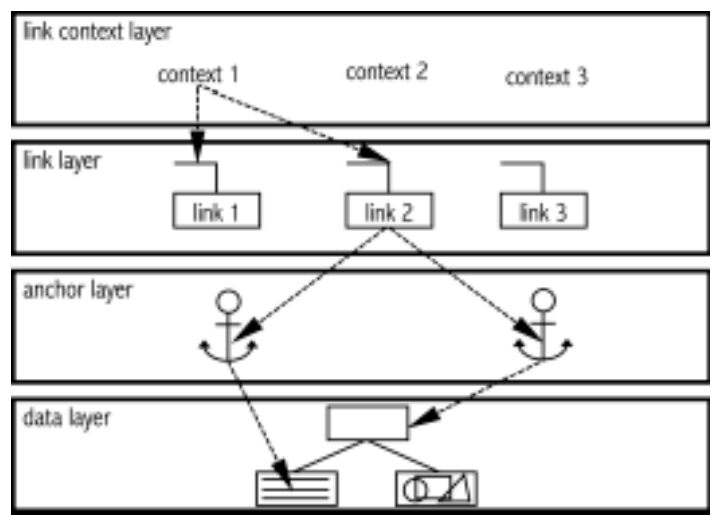

Figure 6: The MIRaCLE link layers

of Xlink/Xpath/Xpointer are incorporated in this model.

The MIRaCLE concept is built upon three layers of linking organisation as shown in figure 6. Each layer is equipped with persistent data describing its state and a communication interface to each of its neighbouring layers. The *anchor* layer carries responsibility for suballocating content within data components. Anchors can dynamically select fragments such as 'all keyword components'. Cells in our content organisation provide a generic anchor interface for addressing its atomic pieces of information.

The *link* layer takes responsibility for relating two or several anchors. Links are set on grounds of fixed references or dynamic selection of anchors in accordance to anchor meta values. A new layer which does not correspond to previous models is created with the *link context*. The link context assigns specific links to a specific view by either a simple selection of link groups or by dynamically extracting links according to semantic or ontological rules. The link context may be set or even defined at runtime and determine the rhetoric of linking as presented to a current user.

## 5   Conclusions and Outlook

Having started from an analysis of needs for educational content management we arrived at an approach, where content components are composite of small to intermediate entities named cells. In this paper we presented solutions to the more general problems of composite content, the treatment of context and the authoring. In addition we roughly introduced a framework of interactivity, a dynamical linking derived from contexts.

Much work, however, has to be done in this ongoing project. Abilities and limitations of this model still wait to be explored by implementing actual learning applications. We hope to arrive here with the beginning of next year.

**Author(s):**

Michael Engelhardt
FHTW Berlin, Hochschulrechenzentrum
Treskowallee 8, D-10318 Berlin, Germany
engelh@fhtw-berlin.de

Arne Hildebrand
FHTW Berlin, Hochschulrechenzentrum
Treskowallee 8, D-10318 Berlin, Germany
hilde@fhtw-berlin.de

Andreas Kárpáti
FHTW Berlin, Hochschulrechenzentrum
Treskowallee 8, D-10318 Berlin, Germany
karpati@fhtw-berlin.de

Torsten Rack
FHTW Berlin, Hochschulrechenzentrum
Treskowallee 8, D-10318 Berlin, Germany
rack@fhtw-berlin.de

Thomas Schmidt, Dr.
FHTW Berlin, Hochschulrechenzentrum
Treskowallee 8, D-10318 Berlin, Germany
schmidt@fhtw-berlin.de